\documentclass[preprint,12pt,authoryear]{elsarticle}

\usepackage{amssymb}
\usepackage{amsthm}
\usepackage{amsmath}

\usepackage{algorithm}
\usepackage{algcompatible}
\usepackage{caption}

\newtheorem{thm}{Theorem}[section]
 \newtheorem{cor}[thm]{Corollary}
 \newtheorem{lem}[thm]{Lemma}
 
\newtheorem{rem}[thm]{Remark}
\newproof{pf}{Proof}

\journal{Journal of Symbolic Computation}

\begin{document}

\begin{frontmatter}



\title{Efficient sparse polynomial factoring \\using the Funnel heap}


\author[label1]{Fatima K. Abu Salem\corref{cor}}
\ead{fa21@aub.edu.lb}

\author[label2]{Khalil El-Harake}
\ead{khalile@bu.edu}

\author[label3]{Karl Gemayel}
\ead{karl@gatech.edu}

\cortext[cor]{Corresponding Author}

\address[label1]{Computer Science Department, American University of Beirut,  Beirut, Lebanon       }

\address[label2]{Computer Science Department, Boston University, Boston, U.S.A.}

\address[label3]{School of Computational Science and Engineering, Georgia Institute of Technology, Georgia, U.S.A.}

\begin{abstract}
This work is a comprehensive extension of \cite{AEG15} that investigates the prowess of the Funnel Heap for implementing sums of products in the polytope method for factoring polynomials, when the polynomials are in sparse distributed representation. We exploit that the work and cache complexity of an Insert operation using Funnel Heap can be refined to depend on the rank of the inserted monomial product, where rank corresponds to its lifetime in Funnel Heap. By optimising on the pattern by which insertions and extractions occur during the Hensel lifting phase of the polytope method, we are able to obtain an adaptive Funnel Heap that minimises all of the work, cache, and space complexity of this phase. This, in turn, maximises the chances of having all polynomial arithmetic performed in the innermost levels of the memory hierarchy, and observes {\it nearly optimal} spatial locality. We provide proofs of results introduced in \cite{AEG15} pertaining to properties of Funnel Heap, several of which are of independent worth extending beyond Hensel lifting. Additionally, we conduct a detailed empirical study confirming the superiority of Funnel Heap over the generic Binary Heap once swaps to external memory begin to take place. We support the theoretical analysis of the cache and space complexity in \cite{AEG15} using accounts of cache misses and memory consumption, and compare the run-time results appearing there against adaptive Funnel Heap. We further demonstrate that Funnel Heap is a more efficient merger than the cache oblivious $k$-merger, which fails to achieve its optimal (and amortised) cache complexity when used for performing sums of products. This provides an empirical proof of concept that the overlapping approach for performing sums of products using one global Funnel Heap is more suited than the serialised approach, even when the latter uses the best merging structures available. Our main conclusion is that Funnel Heap will outperform Binary Heap for performing sums of products, whether data fits in in-core memory or not. 

\end{abstract}

\begin{keyword}
Hensel Lifting \sep Newton Polytopes \sep Polynomial Factorisation \sep Cache Oblivious Algorithms and Data Structures \sep Cache complexity \sep Priority Queues \sep Funnel Heap 



\end{keyword}

\end{frontmatter}







\section{Introduction}

Hensel lifting techniques are at the basis of several polynomial factoring algorithms that are fast in practice. The classical algorithms are designed for generic bivariate polynomials over finite fields without reference to sparsity (e.g. \citep{BostanEtAl04,GL01a}). The polytope method of \citep{AGL04} is intended to factor sparse polynomials more efficiently, by exploiting the structure of their Newton polygon. It promises to be significantly fast when the polygon has a few decompositions, and can help factor families of polynomials which possess the same Newton polytope. While the pre-processing stages of the polytope method benefit from the sparsity of the input in reference to its Newton polygon, the Hensel lifting phase that pursues the boundary factorisations does not do so. Our chain of work in \citep{AEG14,AEG15} reveals that the inner workings of Hensel lifting remain oblivious to the sparsity of the input as well as fluctuations in the sparsity of intermediary output, so long as one is designing the Hensel lifting phase using the dense model for polynomial representation. In contrast, the sparse distributed representation considers the problem size to be a function of the number of non-zero terms of the polynomails treated, which captures the fluctuation in sparsity throughout the factorisation process. In \citep{AEG14}, we revised the analysis of the Hensel lifting phase when polynomials are in sparse distributed representation. We derived that the asymptotic
performance in work, space, and cache complexity is critically affected not only by the degree of the input polynomial, but also by the following factors: (i) the sparsity of each polynomial multiplication, and (ii) the sparsity of the resulting polynomial products to be merged into a final summand. We further showed that even with advanced additive (merging) data structures like the cache aware tournament tree or the cache oblivious $k$-merger, the asymptotic performance of the serialised version in all three metrics is still poor. This was a result of the straightforward implementation which performed polynomial products first off, to be followed by sums of those products, a process that we dubbed {\it serialised}. We remedied this by re-engineering the Hensel lifting phase such that sums of polynomial products are computed simultaneously using a MAX priority queue. This generalises the approach of \citep{John74,MP07,MP09,MP11} for a single polynomial multiplication. We derived orders of magnitude reduction in work, space, and cache complexity even against a serialised version that employs many possible enhancements, and succeeded in evading expression swell. Hereafter, we label the serialised and the priority queue versions of Hensel lifting as SER-HL and PQ-HL respectively. More specifically and with regard
s to the latter algorithm, we will denote by PQ-HL$^{\mathcal{B}}$ the version that uses Binary Heap as a priority queue, and by PQ-HL$^{\mathcal{F}}$ the version that uses Funnel Heap instead. Our experiments in \cite{AEG14,AEG15} demonstrate that the polytope method is now able to adapt significantly more efficiently to sparse input when its Newton polygon consists of a few edges, something not to have been observed when employing SER-HL.

In \citep{AEG15}, we shifted to enhancing the overlapping algorithm PQ-HL$^{\mathcal{F}}$. The motivation lies in the fact that Binary Heap is not scalable, which, on a serial machine, is interpreted to say that its performance will deteriorate once data no longer fits in in-core memory, thus restricting the number of non-zero terms that input and intermediary output polynomials are permitted to possess. By performing priority queue operations using optimal cache complexity and in a cache oblivious fashion, Funnel Heap beats Binary Heap at large scale. The fact that Funnel Heap assumes no knowledge of the underlying parameters such as memory level, memory level size, or word length, makes it ideal for applications where polynomial arithmetic is susceptible to fluctuations in sparsity. However, all of those features can also be observed when adopting an alternate cache oblivious priority queue (see for example, \citep{Bucket04,Arge02}). As such, we pursued Funnel Heap for further attributes that can improve on its asymptotic performance, as well as exploit it at small scale, specifically for Hensel lifting. In \citep{AEG15}, we addressed the chaining optimisation, and how Funnel Heap can be tailored to implement it in a highly efficient manner. We exploited that Funnel Heap is able to identify equal order monomials ``for free'' as part of its inner workings whilst it re-organises itself over sufficiently many
updates during one of its special operations known as the ``SWEEP''. By this we were able to eliminate entirely the requirement for searching from the chaining 
process. We designed a batched mode for chaining that gets overlapped with Funnel Heap's mechanism 
for emptying its in-core components. In addition to also managing expression swell and irregularity in sparsity, batched chaining is sensitive to the number of distinct monomials residing in Funnel Heap, as opposed to the number of replicas chained. This allows the overhead due to batched chaining to decrease with increasing replicas. For sufficiently large input size with respect to the cache-line length, and also sufficiently sparse input and intermediary polynomials, batched chaining that is ``search free'' leads to an implementation of Hensel lifting that exhibits optimal cache complexity in the number of replicas found in Funnel Heap, and one that achieves an order of magnitude reduction in space, as well as a reduction in the logarithmic factor in work and cache complexity, when comparing against PQ-HL$^{\mathcal{B}}$ of \citep{AEG14}. We label as FH-HL the enhancement of Hensel lifting using Funnel Heap and batched chaining.

This paper extends all of the above work in garnering the prowess of Funnel Heap. To this end, we incorporate analytical as well as experimental algorithmics techniques as follows:

\begin{itemize}
\item{In Section \ref{proofs}, we provide proofs of results introduced in \citep{AEG15} pertaining to properties of Funnel Heap, several of which are of independent worth extending beyond Hensel lifting. For example, we provide complete proofs for the following:
\begin{itemize}
\item{We establish where the replicas will reside immediately after each insertion into Funnel Heap.}
\item{We determine the number of times one is expected to call SWEEP on each link of Funnel Heap throughout a given sequence of insertions.}
\item{Given an upper bound on the maximum constituency of Funnel Heap at any one point in time across a sequence of operations, we compute the total number of links required by Funnel Heap.}
\item{We establish that the cache complexity by which one performs batched chaining within FH-HL is optimal.}
\end{itemize}
}
\item{In Section \ref{rank}, we exploit that the work and cache complexity of an Insert operation using Funnel Heap can be refined to depend on the rank of the inserted monomial product, where rank corresponds to its lifetime in Funnel Heap. By optimising on the pattern by which insertions and extractions occur during the Hensel lifting phase of the polytope method, we are able to obtain an adaptive Funnel Heap that minimises all of the work, cache, and space complexity of this phase. This, in turn, maximises the chances of having all polynomial arithmetic performed in the innermost levels of the memory hierarchy, and observes {\it nearly optimal} spatial locality. We show that the asymptotic costs of such preprocessing can be embedded in the overall costs to perform Hensel lifting with batched chaining (FH-HL), independently of the amount of minimisation taking place. We call the resulting algorithm FH-RANK.}
\item{In Section \ref{experimental}, we develop the experimental algorithmics component to our work addressing various facets:
\begin{itemize}
\item{We conduct a detailed empirical study confirming the scalability of Funnel Heap over the generic Binary Heap. By simulating out of core behaviour, Funnel Heap is superior once swaps to external memory begin to take place, despite that it performs considerably more work than Binary Heap. This supports the notion that Funnel Heap should be employed even when performing a single polynomial multiplication or division once data grows out of core.}
\item{We support the theoretical analysis of the cache and space complexity in \citep{AEG15} using accounts of cache misses and memory consumption of FH-HL. This can be seen as an extension of \citep{AEG15}, as the performance measures presented there capture only the real execution time.}
\item{We benchmark FH-RANK against several other variants of Hensel lifting, which include PQ-HL$^{\mathcal{B}}$, PQ-HL$^{\mathcal{B}}$ with the chaining method akin to \cite{MP11}, PQ-HL$^{\mathcal{F}}$, and FH-HL. Our empirical account of time, space and cache complexity of FH-RANK confirm the predicted asymptotic analysis in all three metrics.}
\item{We demonstrate that Funnel Heap is a more efficient merger than the cache oblivious $k$-merger, which fails to achieve its optimal (and amortised) cache complexity when used for performing sums of products. We attribute this to the fact that the polynomial streams to be merged during Hensel lifting cannot be guaranteed to be of equal size (as a result of fluctuating sparsity). This provides an empirical proof of concept that the overlapping approach for performing sums of products using one global Funnel Heap is more suited than the serialised approach, even when the latter uses the best merging structures available.}
\end{itemize}
}
\end{itemize}
We now begin with the following section on background literature and results.
\section{Background}
\label{background}

In the remainder of this paper, we will consider that in-core memory is of size $M$. It is organised using cache
lines (disk blocks), respectively, each consisting of $B$
consecutive words. All words in a single line are transferred
together between in-core and out-of-core memory in one round (I/O
operation) referred to as a cache miss (disk block transfer).

\subsection{Funnel Heap:}

Funnel Heap implements Insert and Extract-Max operations in a cache oblivious fashion. For $N$ elements, Funnel Heap can perform these operations using amortised (and optimal) $O(\frac{1}{B} \log_{M/B} \frac{N}{B})$ cache misses \citep{BF02}.

\begin{figure}[h] 
  \begin{center}
  \includegraphics[width=4in]{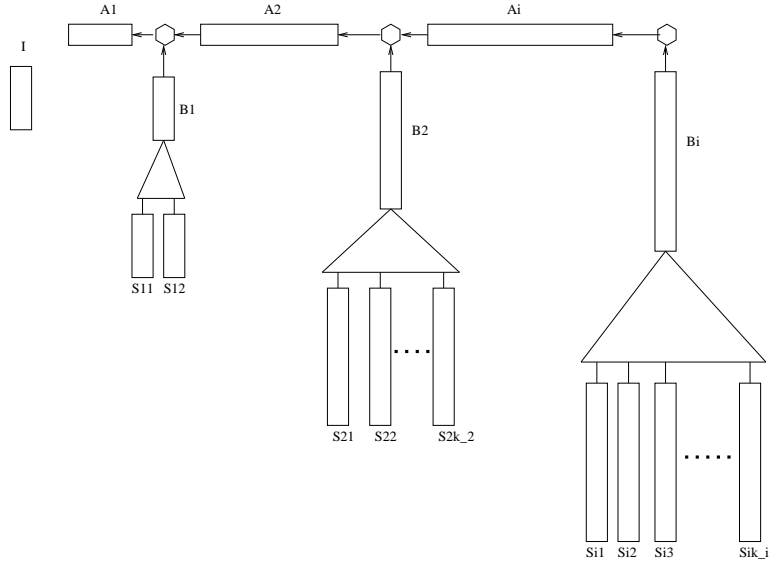}\\
  \caption{Funnel Heap}\label{FunnelHeap}
  \end{center}
\end{figure}

At the innermost level, Funnel heap is first constructed using simple binary mergers. Each binary merger processes two input sorted streams and produces their final merge. The heads of the input streams and the tail of the output stream reside in buffers of a limited size. A binary merger is {\it invoked} using a FILL function when merge steps are repetitively performed until its output buffer is full or both its input streams are exhausted. One can construct binary merge trees by letting the output buffer of one merger be an input buffer of another merger. Now let $k = 2^i$ for $i \in \mathbb{Z}^{+}$. A $k$-merger is a binary merge tree with exactly $k$ input streams. The size of the output
buffer is $k^3$, and the sizes of the remaining buffers are defined
recursively in a Van Emde Boas fashion (See
\citep{BF02a,BF02,FLPR99}). Funnel Heap consists of a sequence
$\{K_i\}$ of $k$-mergers, where $k$ increases doubly exponentially
across the sequence. The $K_i$'s are linked together in a list, with
the help of extra binary mergers and buffers at each juncture of the
list. In Fig. \ref{FunnelHeap}, the
circles are binary mergers, rectangles are buffers, and triangles
are $k$-mergers. Link $i$ in the linked list consists of a binary
merger $v_i$, two buffers $A_i$ and $B_i$, and a merger $K_i$ with
$k_i$ input buffers labeled as $S_{i,1}, \ldots, S_{i,k_i}$. Link $i$
has an associated counter $c_i$ for which $1 \leq c_i \leq k_i+1$.
Initially, $c_i = 1$. It will be an invariant that
$S_{i,c_i},\ldots,S_{i,k_{i}}$ are empty. The first structure in Funnel Heap is a buffer $S_{0,1}$ of
extremely small size $s_1$, dedicated for insertion.
This buffer occupies in-core memory at all times. Funnel Heap is
now laid out in memory in the order $S_{0,1}$, link $1$, link $2$, etc.
Within link $i$ the layout order is $c_i$, $A_i$, $v_i$, $B_i$,
$K_i$, $S_{i,1}$, $\ldots$, $S_{i,k_{i}}$.

The linked list of buffers and mergers constitute one binary tree
$T$ with root $v_1$ and with sorted sequences of elements on the
edges. This tree is heap-ordered: when traversing any path towards
the root, elements will be passed in increasing order. If buffer
$A_1$ is non-empty, the maximum element will reside in $A_1$ or in
$S_{0,1}$. The smaller mergers in Funnel Heap are meant to
occupy primary memory, and can process sufficiently many insertions and extractions
 in-core before an expensive operation is encountered.
In contrast, the larger mergers tend to be out of core, and contain
elements that are least likely to be accessed
in the near future. To perform an Extract-Max, we call FILL on $v_1$ if buffer $A_1$ is
empty. We return the largest element residing in both $S_{0,1}$ and $A_1$.
To insert into Funnel Heap, an element has to be inserted
into $S_{0,1}$. If $S_{0,1}$ is full, a SWEEP function is called. Its purpose is
to free the insertion buffer $S_{0,1}$ together with all the heavily
occupied links in Funnel Heap which are closer to in-core memory.
During a SWEEP, all elements residing in those dense links are
extracted then merged into one single stream. This stream is then
copied sufficiently downwards in Funnel Heap, towards the first link which 
has at least one empty input buffer. As a result of SWEEP,  
the dense links are now free and Funnel Heap operations are resumed within in-core memory. The SWEEP kernel is considerably expensive, yet, sufficiently many insertions and all the extractions can be accounted for between any two SWEEPs. 

\subsection{The polytope method:}


Let $\mathbb{F}$ denote a finite field of characteristic $p$, and
consider a polynomial $f \in \mathbb{F}[x,y]$ with total degree $n$.
We wish to obtain a polynomial factorisation of $f$ into two factors
$g$ and $h$ such that $f = gh$ and $g$, $h \in {\mathbb F}[x,y]$.
 Let $Newt(f)$ denote the Newton polygon $\mathbb{R}^2$ of $f$ defined as the convex hull of the support
vector of $f$. One identifies suitable subsets $\{\Delta_i\}$ of
edges belonging to $Newt(f)$, such that all lattice
points can be accounted for by a proper translation of this set
of edges. One then specialises terms of $f$ along
each edge $\delta_j^{(i)} \in \Delta_i$. Those specialisations are
derived from the nonzero terms of $f$ whose exponents make up
integral points on each $\delta_j^{(i)}$, and we label them
as $f_{0}^{\delta_{j}}$. These can be transformed into Laurent polynomials in one variable. For at
least one $\Delta_i$, the associated edge polynomials
$f_{0}^{\delta_{j}}$ ought to be squarefree, for all $\delta_j \in
\Delta_i$. One then begins lifting using the boundary
factorisations given by $f_0^{\delta_{j}} =
g_{0}^{\delta_{j}}h_{0}^{\delta_{j}}$, for all $\delta_j \in
\Delta_i$. For each boundary factorisation, we determine
the associated $\{g_k\}$'s and $\{h_k\}$'s that satisfy the Hensel
lifting equation
\begin{equation}
g_{0}^{\delta_j} h_{k}^{\delta_j} + h_{0}^{\delta_j}
g_{k}^{\delta_j} = f_{k}^{\delta_j} - \sum_{j = 1}^{k-1}
g_{j}^{\delta_j} h_{k-j}^{\delta_j} \label{maineq2}
\end{equation}
for $k = 1,\ldots, \min(\deg(g_0), \deg(h_0))$.

\subsection{Sums of products using a priority queue}

In \citep{AEG14} we revised the analysis associated with the
bottleneck in computation arising in Eq. (\ref{maineq2}), using the
sparse distributed representation. In this model of representation, a polynomial is exclusively represented as the sum of its non-zero terms, sorted upon 
some decreasing monomial ordering. 
Eq. (\ref{maineq2}) can be
modeled using the input and output requirements shown in Alg. 1:

\begin{algorithm}
\label{local-iter} \caption{Local-Iterative}
\begin{algorithmic}[1]\REQUIRE An integer $k$ designating one iterative step in the Hensel lifting process. Two sets of univariate polynomials over $\mathbb{F}$, $\{g_i\}_{i = 1}^{k-1}$, $\{h_i\}_{i = 1}^{k-1}$, in sparse distributed monomial order representation.
\ENSURE The polynomial $S_{k} = \sum_{i = 1}^{k-1} g_{i} \cdot h_{j}$, where $j = k-i$.
\medskip
\FOR{$i = 1$ to $k-1$} \STATE Compute $p_{i} \leftarrow g_{i} \cdot
h_{j}$. \ENDFOR \STATE Compute $S_k = \sum_{i = 1}^{k-1} p_i$.
\end{algorithmic}
\end{algorithm}

We distinguish between the serialised approach (SER-HL) and the overlapping approach (PQ-HL) for performing the required arithmetic. In the serialised
version, one performs all polynomial multiplications first, and then merges all the resulting polynomial products. In the overlapping approach, one handles all arithmetic
simultaneously using a single Max priority queue. 
In \citep{AEG14}, we analysed the work, space, and cache complexity, when polynomials are 
in sparse distributed representation. We derived that the performance of the serialised version 
in all three metrics is critically affected not only by the degree of the input polynomial,
but also by the following factors: (i) the sparsity of each
polynomial multiplication, and (ii) the sparsity of the resulting
polynomial products to be merged into a final summand. We further
showed that this remains the case even with advanced additive (merging) data structures
like the cache aware tournament tree or the cache oblivious
$k$-merger, for performing the sums of resulting polynomial products, and that the serialised approach is not
able to fully exploit the cache efficiency of these structures.

In the overlapping approach, the priority queue is initialised using the highest order monomial products generated 
from each product $g_i \cdot h_j$. Then, terms of $S_k$ are
produced in decreasing order of degree, via successive invocations of Extract-Max upon the priority queue. 
In \citep{AEG15}, we pursued Funnel Heap as an alternative to the generic Binary Heap for implementing the overlapping approach. Beyond its cache oblivious nature and optimal cache complexity, we showed that Funnel Heap allows for a mechanism of chaining that significantly improves its overall performance. Chaining replicas outside the priority queue following insertions is a well known technique (e.g. see \citep{MP07,MP09,MP11}) for the case of single polynomial multiplication using binary heap). It helps reduce several parameters tied to performance, such as the total number of extractions required to perform a single polynomial multiplication and the size of the priority queue. In turn, the latter results in reducing the number of monomial comparisons as well as the cache complexity required to perform each priority queue operation. In the straight-forward implementation, one has to search for a replica immediately after an insertion and then chain the newly inserted element to the end of a linked list tied to that replica in the priority queue. When using Binary Heap, chaining hinders performance critically. Each insertion into the linked list denoting the chain incurs a random miss, whereas a single search query may require traversing the entire heap. It follows that the work and cache complexity of a single insertion amounts to that of traversal of $N$ elements for a heap of size $N$. When employed in the priority queue that is implementing sums of products arising in Hensel lifting, chaining becomes daunting as the size of the queue and the amount of replication change irregularly from one iteration to the other.

In \citep{AEG15} we showed how to exploit the expensive SWEEP kernel of Funnel Heap in order to develop a cache friendly batched chaining mechanism (BATCHED-CHAIN) that gets intertwined with the SWEEP's internal operations. The crux behind our approach lies in delaying chaining and performing it in batches, somehow at the ``right time''. In the interim, a prescribed amount of replication is tolerated, whose effect is shown to be insignificant at scale. Here, we restrict chaining to only two specific phases in Funnel Heap's operations. If one is inserting a monomial product into the (sorted) insertion buffer $S_{0,1}$, a replica that resides in $S_{0,1}$ is immediately identified and chaining can take place. One does not attempt to find a replica outside of $S_{0,1}$. If such a replica exists, chaining will be deferred until $S_{0,1}$ is full. That is when SWEEP is invoked upon some link $i$ as well as one of its input buffers $S_{i,c_i}$. In the duration of SWEEP, one is forming the stream $\sigma$ which contains the merged output of all elements in the buffers leading from $A_i$ to $S_{i,c_i}$ together with all elements in links $1, \ldots, i-1$. During the merge, the replicas residing in those specified regions of Funnel Heap will be aligned consecutively and thus identified. One can then chain them all and at once outside of Funnel Heap. 

BATCHED-CHAIN eliminates entirely the need for searching for replicas, and lesser links would be allocated to Funnel Heap, which reduces garbage collection. BATCHED-CHAIN is further sensitive to the number of distinct monomials in Funnel Heap, and not the number of replicas chained. This can be understood to mean that the overhead due to chaining decreases with increasing replicas, which is intuitively appealing, since chaining is likely to be disabled once the number of replicas is lower than an acceptable threshold. When incorporating Funnel Heap and BATCHED-CHAIN into the priority queue algorithm for sums of products, Alg. FH-HL was shown to be significantly fast. The timings reported in \citep{AEG15} correspond to overall run-time, with the following percentages of improvement recorded, attained with increasing input size: about 90\%-98\% (FH-HL to Magma 2.18-7), about 90\%- 99\% (FH-HL to SER-HL), about 10\%-60\% (FH-HL to PQ-HL$^{\mathcal{B}}$). The dramatic reduction in run-time over SER-HL is largely attributed to substantial expression swell, and that over PQ-HL$^{\mathcal{B}}$ is attributed to BATCHED-CHAIN.


\section{Funnel Heap Properties: Extended Results}
\label{proofs}

In this section we revisit several claims made in \citep{AEG15} and provide their complete proofs. Those results pertain to the behaviour of Funnel Heap in general and not necessarily only in relation to Hensel lifting, and thus are of independent worth. Unless otherwise stated, all lemmas and corollaries in this specific section are stated in \cite{AEG15}.

We begin by the following invariant which identifies where the replicas will reside immediately after each Insert into Funnel HEap:
\begin{lem}
Let $\ell$ denote the index of the last link in Funnel Heap. Using BATCHED-CHAIN, and immediately following each insertion, there will be no replication within the constituency of any buffer $\{\{S_{i,j}\}_{j = 1}^{k_{i}}\}_{i = 0}^{\ell}$. As a result, a given element in some buffer $S_{i,j}$ may only be replicated at most once in each of the preceding buffers $\{S_{i,j'}\}_{j' = 1}^{j-1}$ in its own link or in each of the buffers $\{\{S_{i',j}\}_{j = 1}^{k_{i'}}\}_{i' = i+1}^{\ell}$ in the larger links.
\label{noreplicas}
\end{lem}
\begin{pf}
Consider the case when one is inserting immediately into the insertion buffer $S_{0,1}$. Alg. BATCHED-CHAIN ensures that chaining is happening immediately, and so there will be no replicas in this particular buffer. Now consider a random $S_{i,j}$ for $i > 0$. We know that one can only write elements to $S_{i,j}$ upon a call onto $SWEEP(i)$. This call produces the stream $\sigma$ which merges the content of all links $1,\ldots,i-1$ together with the content of the path $p$ leading from $A_1$ down to $S_{i,j}$. Since BATCHED-CHAIN employs chaining during the formation of $\sigma$, buffer $S_{i,j}$ will not contain any replicas. 

Now, by the first claim above, each buffer $S_{i,j}$ in Funnel Heap contains distinct elements. When $i = 0$, it is straightforward to see that since $S_{0,1}$ has no buffers which precede it, each of its elements is replicated at most once in each of the following buffers. Now take $i > 0$. We know that once SWEEP is called onto $S_{i,j}$, each buffer $\{S_{i,j'}\}_{j' > j}$ in the $i$'th link must be empty. Also, as we form $\sigma$ -- the end of which is written to $S_{i,j}$ -- we exclude the elements residing in each buffer that is also in the same link as $S_{i,j}$ but which precede it in that link. It follows that the only possible replicas of each element in $S_{i,j}$ will be in each of the buffers $\{S_{i,j'}\}_{j' = 1}^{j-1}$ preceding it in its own link, as well as each of the buffers $\{\{S_{i',j}\}_{j = 1}^{k_i}\}_{i' = i}^{\ell}$ in the larger links. 
\end{pf}

The following result captures the number of times one is expected to call SWEEP on each link of Funnel Hap throughout a given sequence of insertions and extractions:
\begin{lem}
Let $\ell$ denote the index of the last link in Funnel Heap and let $T_j$ denote the total number of times SWEEP$(j)$ is called, across a given sequence of insertions and extractions. Then 
\[
T_j = c_{\ell}\cdot \overset{\ell-1}{\underset{i = j}\prod }  k_i
\]
\label{totalsweeps}
\end{lem}
\begin{pf}
We proceed by backward induction on $j$. Take $j = \ell$. Link $\ell$ has $k_{\ell}$ input buffers. Since this is the last link, not all of its input buffers $S_{i,j}$ may be written onto using SWEEP. In fact, exaclty $c_{\ell}$ of them will be so. We thus have $T_{\ell} = c_{\ell}$. We now show that $T_j = c_{\ell} \cdot \overset{\ell-1}{\underset{i = j}\prod }  k_i$ assuming the property holds for $T_{j+1}$. Observe that before any SWEEP on link $j+1$ has occurred, there should have preceded it exactly $k_{j}$ SWEEPs, in order to fill each of the input buffers in link $j$. Also, by the inductive hypothesis, the total number of SWEEPs on link $j+1$ is given by $T_{j+1} = c_{\ell}\overset{\ell-1}{\underset{i = j+1}\prod }  k_i$. Combining, we get that there are
\[
\begin{array}{rcl}
T_j & = & k_j \cdot c_{\ell} \cdot \overset{\ell}{\underset{i = j+1}\prod }  k_i \\
& = & c_{\ell} \cdot \overset{\ell}{\underset{i = j}\prod }  k_i\\
\end{array}
\]
SWEEPs on link $j$.
\end{pf}

Given an upper bound on the maximum constituency of Funnel Heap at any one point in time across a sequence of operations, we now determine the total number of links the heap requires:
\begin{lem}
Let $\ell$ denote the index of the last link in Funnel Heap. Then 
\[
\ell = \theta(\left \vert T \right \vert  \log \log \left \vert T \right \vert),\]
where $\left \vert T \right \vert$ designates the maximum number of elements residing in Funnel Heap at any point in time.
\label{last-versus-total}
\end{lem}
\begin{pf}
From \citep{BF02} we invoke the following proven results which we require for our proof:
\begin{enumerate}
\item{The space usage $s_i$ of each input buffer in link $i$ satisfies $s_i = \theta(k_i^3)$, where $k_i$ is the number of input buffers in link $i$.}
\item{The space usage of link $i$ is $\theta(k_is_i)$, i.e. it is dominated by the space usage of all of its $k_i$ input buffers.}
\item{$k_i = \theta(k_{i-1}^{4/3})$}
\end{enumerate}
Since link $\ell$ is the last link required by Funnel Heap to host all elements of its elements, those elements will consume at least one path leading to the first input buffer of link $\ell$, and at most all $k_{\ell}$ such possible paths. By (2) above, the space usage of each such path is dominated by the size of the input buffer itself and we thus have $\left \vert T \right \vert = O(k_{\ell}s_{\ell})$ 
and $\left \vert T \right \vert = \Omega(s_{\ell})$. By $T = O(k_{\ell}s_{\ell})$ we have:
\begin{equation*}
\begin{array}{rcl}
T & = &  O(k_{\ell}s_{\ell})\\
& = & O(k_{\ell}^{4}) \quad \mbox{ by (1) above}\\
& = & O\left (\left (k_{1}^{(4/3)^{\ell-1}}\right)^{4}\right) 
\end{array}
\end{equation*}
where the last equality follows by (3) above and by unrolling the recursive relation down to the base case. Using $k_1 = 2$ and composing the logarithm function on the two bases 2 and 4/3 respectively, we get 
$\ell = O(\log \log T)$.

Taking $T = \Omega(s_{\ell})$ one can proceed analogously as above and obtain $\ell = \Omega(\log \log T)$. This concludes the proof.
\end{pf}

As in \cite{MP07,MP09,MP11}, reasoning in the sparse distributed representation produces worst-case versus best case polynomial multiplication, depending on the structure of the output. In the worst case, a given multiplication $g_i\cdot h_j$ is sparse as it yields a product with $\theta(\#g_i \cdot \#h_j)$ non-zero terms, an incidence of a memory bound computation. At best, the multiplication is dense as it yields a product with $\theta(\#g_i + \#h_j)$ terms. When the product has significantly fewer terms due to cancelation of terms, the operation is said to suffer from expression swell. We now establish that the cache complexity by which one performs BATCHED-CHAIN within FH-HL is optimal. For this, we require a few notations from \cite{AEG15} that will be helpful in the forthcoming sections as well. Let $\bar{g} = \max\{\#g_i\}_{i = 1}^{k}$ and $\bar{h} = \max \{\#h_j\}_{j = 1}^{k}$, which denote the maximum number of non-zero monomials comprising each $g_j$ and $h_j$ respectively. Let $\tau$ denote the fraction of reduction in the size of the heap during chaining, such that the largest size the priority queue attains during the $k$'th lifting step is $\theta(k \bar{g}/\tau)$. Let $\tau'$ denote the fraction of replication in the total number of monomial products such that the total number of replicas chained during the $k$'th Hensel lifting step is $\theta(k\bar{g}\bar{h}/\tau')$. The two parameters $\tau$ and $\tau'$ reflect, in an asymptotic sense, the changes in the size of the queue as a function of the amount of replicas. Particularly, the bounds on $\tau$ and $\tau'$ are as follows. When no replicas are encountered at all during any one lifting step, we have that $\tau = 1$ and $\tau' = \theta(k \bar{g}\bar{h})$. In contrast, when each polynomial in the pair $(g_i,h_j)$ is totally dense and all resulting products in one lifting step are of the same degree, the heap will contain only one element, leading to $\tau = \theta(k \bar{g})$ and $\tau' = \theta(1)$.

We now have the following:
\begin{cor}
Assume the sparse distributed representation for polynomials. Assume further that $B = O \left ( \frac{\tau \bar{h} \log (k \bar{g}/\tau)}{\log \log (k\bar{g}/\tau)}\right)$. In the worst case analysis when each polynomial multiplication $g_ih_j$ is sparse, the cache complexity by which one performs BATCHED-CHAIN within FH-HL is optimal.
\label{cor1}
\end{cor}
\begin{pf}
Following the analysis in Prop. 3.6 of \cite{AEG15}, the cache complexity of FH-HL is split into two major parts. The first part accounts for all the insertions into Funnel Heap using $O(k\bar{g}\bar{h} \frac{1}{B} \log_{M/B} \frac{k\bar{g}}{\tau})$ cache misses. The second part accounts for the cost to perform BATCHED-CHAIN using $O(\frac{k \bar{g}\bar{h}}{\tau'\, B} + \frac{k \bar{g}}{\tau} \log \log \frac{k \bar{g}}{\tau})$ cache misses. When $B = O \left ( \frac{\tau \bar{h} \log (k \bar{g}/\tau)}{\log \log (k\bar{g}/\tau)}\right)$, we get that the second summand in the cache complexity incurred by BATCHED-CHAIN is dominated by the cost to perform all the insertions into Funnel Heap, or that the cost for BATCHED-CHAIN is dominated by $O(\frac{k \bar{g}\bar{h}}{\tau'\, B})$, where $\theta(\frac{k \bar{g}\bar{h}}{\tau'})$ denotes the total number of replicas chained. It follows that the cache complexity of BATCHED-CHAIN corresponds to that of traversal, and hence is optimal. 
\end{pf}

In the following, we provide a detailed proof that FH-HL, and thanks to BATCHED-CHAIN, outperforms PQ-HL$^{\mathcal{F}}$ (and thus by transitivity, also PQ-HL$^{\mathcal{B}}$). In other words, performing sums of products using Funnel Heap with BATCHED-CHAIN is provably more efficient in work, space, and cache complexity than if we were to resort to a standalone Funnel Heap implementation.
\begin{cor}
Assume the sparse distributed representation for polynomials, and assume further the conditions in Cor. \ref{cor1}. In the worst case analysis when each polynomial multiplication $g_ih_j$ is sparse, FH-HL achieves an order of magnitude reduction in space, as well as a reduction in the logarithmic factor in work and cache complexity, over PQ-HL$^{\mathcal{F}}$.
\label{cor2}
\end{cor}
\begin{pf}
From \citep{AEG14}, Alg. PQ-HL$^{\mathcal{F}}$ requires the following costs:
\begin{table}[!htbp]
\begin{center}
\begin{tabular}{|c|c|c|} \hline
Space & Work & cache complexity   \\\hline
$\theta(k\bar{g})$ & $\theta(k \bar{g}\bar{h} \, \log k \bar{g})$ & $O \left(k \bar{g}\bar{h} \frac{1}{B} \log_{M/B} k\bar{g}\right)$  \\\hline
\end{tabular}
\end{center}
\end{table}

From \citep{AEG15}, Alg. FH-HL requires the following costs:
\begin{table}[!htbp]
\begin{center}
\begin{tabular}{|c|c|c|} \hline
Space & Work & cache complexity   \\\hline
$\theta\left (\frac{k \bar{g}}{\tau}\right)$ & $\theta \left (k \bar{g}\bar{h} \log \frac{k \bar{g}}{\tau}\right)$ & $O \left (k \bar{g} \bar{h} \frac{1}{B} \log_{M/B} \frac{k\bar{g}}{\tau}  + \frac{k\bar{g}\bar{h}}{\tau'B} \right)$ \\\hline
\end{tabular}
\end{center}
\end{table}

Reductions in space borne by FH-HL are obvious by comparing $\theta\left (\frac{k \bar{g}}{\tau}\right)$ and $\theta \left(k \bar{g}\right)$ respectively, and noting that $\tau \geq 1$. The logarithmic factor reductions in work are obvious by comparing $\theta \left (k \bar{g}\bar{h} \log \frac{k \bar{g}}{\tau}\right)$ and $\theta \left(k \bar{g}\bar{h} \log k \bar{g}\right)$ respectively. Similarly, for reductions in cache complexity, we require 
\begin{equation*}
\left (k \bar{g} \bar{h} \frac{1}{B} \log_{M/B} \frac{k\bar{g}}{\tau}
\right) + \frac{k\bar{g}\bar{h}}{\tau'B}= O \left (k \bar{g} \bar{h}
\frac{1}{B} \log_{M/B} k\bar{g}\right)
\end{equation*}
or
\begin{equation}
\left (\frac{1}{B} \log_{M/B} \frac{k\bar{g}}{\tau}
\right) + \frac{1}{\tau'B}= O \left (\frac{1}{B} \log_{M/B} k\bar{g}\right)
\label{req}
\end{equation}
Recall the bounds established earlier for $\tau$ and $\tau'$. When $\tau = 1$, we have $\tau' = \theta \left (k \bar{g}\bar{h}\right)$, 
for which we have:
\[
\left (\frac{1}{B} \log_{M/B} \frac{k\bar{g}}{\tau}
\right) + \frac{1}{\tau'B} = \theta \left ( \frac{1}{B} \log k \bar{g} + \frac{1}{k \bar{g}\bar{h} \, B}\right) = \theta \left ( \frac{1}{B} \log k \bar{g}\right)
\]
and (\ref{req}) holds. As $\tau$ increases, $\tau'$ satisfies $\tau' = \Omega(1)$ and so
\[
\left (\frac{1}{B} \log_{M/B} \frac{k\bar{g}}{\tau}
\right) = O\left (\frac{1}{B} \log_{M/B} k\bar{g} \right)
\]
and 
\[
\frac{1}{\tau' \, B} = O \left ( \frac{1}{B}\right) = O\left (\frac{1}{B} \log_{M/B} k\bar{g} \right)
\]
for which (\ref{req}) holds again. This concludes the proof. 
\end{pf}

\section{Adaptive Funnel Heap and the Sequence of Insertions/Extractions}
\label{rank}

The canonical SWEEP function described in Sec. \ref{background} works by identifying the smallest link in Funnel Heap that is completely empty, which necessitates that one keeps pushing the content of the heap downwards in the direction of larger and larger links, which are also more likely to be out-of-core. An enhanced version of SWEEP exploits the smaller links in Funnel Heap that are sufficiently sparse, instead of always sweeping onto totally empty, yet significantly larger buffers. The refined SWEEP operation identifies the first link $i$ whose total number of elements residing in its input buffers $\{S_{i,j}\}$ is less than half of its total size. The input buffer with minimal occupancy in that link, say $S_{i,j_1}$, is then recycled and its content moved onto another input buffer, say $S_{i,j_2}$, with second largest occupancy. SWEEP is now called with $S_{i,j_1}$ as the destination buffer. That smaller buffers are effectively used instead of the larger, out-of-core buffers causes Funnel Heap to adapt to various modes of usage. The analysis in \citep{BF02} shows that the amortised cost for the $r$'th insertion is now $O(\frac{1}{B} \log_{M/B} \frac{N_r}{B})$, where $N_r$ denotes some notion of the lifetime of the $r$'th inserted element in Funnel Heap. Particularly, if the $r$'th inserted element is removed by an Extract-Max prior to the $t$'th inserted element, then $N_r = t-r$.

In this section, we exploit the idea that the cache complexity of an Insert operation can be refined to depend on the rank of the inserted monomial product, by optimising on the pattern by which insertions and extractions occur during the Hensel lifting phase, with the notion of lifetime in hindsight. We achieve this by efficiently delaying all the insertions that come from polynomial pairs that ``can wait'', as indicated by their total order and their rank in relation to the maximal element residing in Funnel Heap. The refined process, which we label as FH-RANK, proceeds as follows. We first accumulate the set of all distinct monomial orders $\alpha$ appearing in the sum of products $S_k$. When the input polynomials are univariate, the monomial order can be understood to denote the total degree of a given monomial product. For each $\alpha$, let $\psi(\alpha)$ denote the set of indices $\{i \left \vert \right. 1 \leq i \leq k \wedge o(g_ih_{j=k-i}) = \alpha\}$, which maps each monomial order $\alpha$ to the polynomial operands that resulted in a product of this particular order. Let $\mathcal{O} = \{(\alpha,\psi(\alpha))\}$, where $\mathcal{O}$ is sorted on $\alpha$ in strictly decreasing order. We manipulate the sequence of insertions into Funnel Heap based on information derived from $\mathcal{O}$ as follows. Let $\alpha_{\max} = \max \{\alpha\}$. Initially, we insert monomial products generated from pairs pointed to by $\psi(\alpha_{\max})$ only. No other polynomial pair of total order $\alpha' < \alpha_{\max}$ may be involved, until at least one monomial product from $\psi(\alpha_{\max})$ has been inserted into Funnel Heap, whose order is less than or equal to $\alpha'$. This point in time is identified by knowledge of the next maximal order to be encountered before an upcoming Extract-Max is called. The function NEXT-MAX-ORDER introduced below answers this particular query, by calling EXTRACT-MAX on funnel heap whilst refraining from actually extracting the maximal element and only reporting on its order. 

Alg. 2 summarises the details of this adaptive technique, which we label as FH-RANK. We further demonstrate its impact asymptotically speaking, particularly with regards to the costs associated with preprocessing the sorted list $\mathcal{O}$, as well as invoking NEXT-MAX-ORDER on top of the existing calls to Extract-Max. 

\newpage

\begin{algorithm}
\caption{Alg. FH-RANK}\label{global-fh-minrank}
\begin{algorithmic}[1]
\REQUIRE An integer $k$ designating one iterative step in the Hensel lifting process. Two sets of univariate polynomials over $\mathbb{F}$, $\{g_i\}_{i = 1}^{k-1}$, $\{h_i\}_{i = 1}^{k-1}$, in sparse and sorted monomial order representation. Also, two arrays $Ord_g$ and $Ord_h$, such that $Ord_g(i)$ and $Ord_h(i)$ designate respectively the maximal order of the polynomials $g_i$ and $h_i$ under the assumed monomial ordering, for $i = 1,\ldots,k$.
\ENSURE The polynomial $S_{k} = \sum_{i = 1}^{k-1} g_{i} \cdot h_{j}$, where $j = k-i$.
\bigskip
\STATE For each product pair $(g_i,h_j)$, calculate $o(i,j)$, the total order of their product under the assumed monomial ordering, by a forward scan of the array $Ord_g$ and a backward scan of $Ord_h$. Collect the set $\{\alpha\}$ of distinct total orders. 
\STATE Set $\mathcal{O} = \{\left (\alpha, \psi(\alpha)\right)\}$, where $\psi(\alpha)= \{i \left \vert \right. 1 \leq i \leq k \wedge o(g_ih_{j=k-i}) = \alpha\}$. If $k \in \theta(n)$, sort $\mathcal{O}$ on the $\{\alpha\}$'s using Counting Sort. Else, use a cache efficient comparison based algorithm.
\STATE Consider $O_{ind} = (\alpha_{ind}, \psi(\alpha_{ind}))$, where $ind \leftarrow 1$. 
\FOR {$i \in \psi(\alpha_{ind})$, and $j = k-i$} 
 \STATE Call BATCHED-CHAIN$(X_{1}^{(i)}Y_{1}^{(j)}, {\bf g}^{(i)}, {\bf h}^{(j)})$ to insert those monomial products into Funnel Heap while chaining.
\ENDFOR 
\STATE Set $t \leftarrow 0$. \REPEAT \STATE Let $(XY, {\bf g}, {\bf h})$ denote the the maximal element in Funnel Heap, $\beta$ denote the rank of $XY$ under
the assumed monomial ordering. Set $t \leftarrow t+1$, $a_{t} \leftarrow 0$, $R_t \leftarrow XY$. 
\STATE Call
{\it Extract-Max} on Funnel Heap to return the maximal element $(XY, {\bf g}, {\bf h})$. 
\STATE Return all monomial products of order $\beta$ chained outside of Funnel Heap. 
\WHILE {the
maximal element in Funnel Heap has rank equal to $\beta$} \STATE Repeat
Steps 10-11 above \ENDWHILE
\FOR {each element $(X_{u}^{(i)}Y_{w}^{(j)}, {\bf g}^{(i)}, {\bf h}^{(j)})$ returned in Steps 10 and 11}
\STATE Perform the coefficient arithmetic required to accumulate in $a_t$ by
reading the coefficients of terms pointed to by ${\bf g}^{(i)}$ and ${\bf
h}^{(j)}$, then set $S_k \leftarrow S_k + a_tR_t$.
\algstore{myalg}
\end{algorithmic}
\end{algorithm}

\begin{algorithm}
\ContinuedFloat
\caption{Alg. FH-RANK (continued)}
\begin{algorithmic}
\algrestore{myalg}
\STATE If $w < \# h_{j}$, insert into Funnel Heap the horizontal successor
by calling BATCHED-CHAIN$(X_{u}^{(i)}Y_{w+1}^{(j)})$.
\STATE If $w = 1$ and $u < \# g_{i}$, insert into Funnel Heap the vertical successor
by calling BATCHED-CHAIN$(X_{u+1}^{(i)}Y_{w}^{(j)})$.
\ENDFOR
\STATE $\beta' \leftarrow NEXT-MAX-ORDER$.
\WHILE {$\alpha_{ind+1} \geq \beta'$}
\FOR {$i \in \psi(\alpha_{ind+1})$, and $j = k-i$}
 \STATE Call BATCHED-CHAIN$(X_{1}^{(i)}Y_{1}^{(j)}, {\bf g}^{(i)}, {\bf h}^{(j)})$ to insert those monomial products into Funnel Heap while chaining.
 \STATE $ind \leftarrow ind+1$.
\ENDFOR
\ENDWHILE
\UNTIL{no monomials can be inserted into Funnel Heap.} \STATE Return $S_k$.
\end{algorithmic}
\end{algorithm}


%
%

Lem. \ref{lifetime}, Cor. \ref{IO-min} and Cor.\ref{queue-minimised} we argue that with FH-RANK, the space required to handle sums of products using the priority queue approach is minimised, and so are the work and cache complexity required to perform insertions: 
\begin{lem}
In Alg. FH-RANK, the lifetime of each inserted element in the priority queue is minimised.
\label{lifetime}
\end{lem}
\begin{pf}
We establish the proof by showing that no monomial product enters the priority queue prior to the time when it is necessary for it to be there. Put differently, a monomial product is inserted into Funnel Heap at a point in time when its insertion can no longer be deferred. To show the claim, assume that Funnel Heap contains a monomial product $(X_{u}^{(i)}Y_{w}^{(j)}, {\bf g}^{(i)}, {\bf h}^{(j)})$ whose insertion could have been safely deferred. Then one of those two cases must hold:
\begin{itemize}
\item{The polynomial pair $(g_i,h_j)$ should have not been engaged in the insertion process. But that is impossible since the pair $(g_i,h_j)$ must have been identified by the latest call to NEXT-MAX-ORDER in Step 20, which ensures that a polynomial pair is chosen only when its highest order monomial product is larger than the maximum residing element in Funnel Heap.}
\item{The pair $(g_i,h_j)$ is already engaged in the insertion process but this particular monomial product $X_{u}^{(i)}Y_{w}^{(j)}$ can wait. This is also impossible since the sequence of insertions and extractions ensures that a given monomial product is inserted only after one of its horizontal or vertical predecessors ($X_{u-1}^{(i)}Y_{w}^{(j)}$ or $X_{u}^{(i)}Y_{w-1}^{(j)}$) have been extracted. This means that ($X_{u}^{(i)}Y_{w}^{(j)}$) can potentially be the next maximum, and so must be inserted into Funnel Heap.}
\end{itemize}
\end{pf}
\begin{cor}
In Alg. FH-RANK, the cache complexity of Insert is minimised. \label{IO-min}
\end{cor}
\begin{pf}
Consider the $r$'th insertion in the sequence prescribed by Alg. FH-RANK. Let $t$ be the index of the first insertion that takes place immediately following the extraction of the $r$'th element. The refined SWEEP operation attains the amortised cache complexity of the $r$'th insertion to be $O(\frac{1}{B} \log_{M/B} \frac{N_r}{B})$, where $N_r = t-r$. By Prop. \ref{queue-minimised}, the lifetime in Funnel Heap of the $r$'th element is minimised, and hence, so is $t-r$. This concludes the proof. 
\end{pf}
Moreover, we have:
\begin{cor}
In Alg. FH-RANK, the size of the priority queue is minimised. Put differently, the likelihood that it can operate in as innermost as possible levels of the memory hierarchy is maximised. \label{queue-minimised}
\end{cor}
\begin{pf}
The size of the priority queue is minimised as an immediate consequence of Lemma \ref{lifetime}, since no element enters the queue prior to the time when it has to be there. 
\end{pf}
As an immediate consequence of Cor. \ref{queue-minimised}, the work required to perform each monomial product using insertions and extractions is also minimised.

Finally, we establish that spatial locality associated with BATCHED-CHAIN in Alg. FH-RANK is {\it nearly optimal}. By observing {\it optimal spatial locality}, the length of each stride in the address space is at most $1$. Our definition of {\it nearly optimal} relaxes this requirement: it suffices to have that the length of each stride in the address space is minimised.
\begin{cor}
BATCHED-CHAIN invoked by Alg. FH-RANK exhibits nearly optimal spatial locality. 
\end{cor}
\begin{pf}
From \cite{AEG15}, the mechanism for chaining in batches introduced in Sec. \ref{background} is achieved as follows. We store all monomials of a given order $\alpha$ and that have to be excluded from the queue in a dynamic array $D[\alpha]$. The set $\{D[\alpha]\}_{\alpha}$ over all monomial orders $\alpha$ encountered during the latest SWEEP represents pointers to the heads of chains. These pointers are aligned consecutively in a static array $D$ in increasing monomial order, where the size of $D$ grows like the bound on $\deg(S_k)$. We will label the memory accesses to the dynamic array pointed to by $D[\alpha]$ as horizontal accesses. In contrast, we will label the memory accesses to the static array $D$, as we hop from one pointer $D[\alpha]$ to another, as vertical accesses. Observe that replicas of each given monomial order $\alpha$ are chained consecutively into the single chain pointed to by $D[\alpha]$, which maintains sequential spatial locality corresponding to strides of length equal to $1$ in the horizontal direction. Because of BATCHED-CHAIN, all pointers $\{D[\alpha]\}_{\alpha}$ are accessed in increasing monomial order. Additionally, because of FH-RANK, no element being chained could have been delayed entry into Funnel Heap. It follows that jumps in the vertical direction are minimised.
\end{pf}

We devote the remainder of this section to showing that the pre-processing costs associated with Alg. FH-RANK can be embedded in the costs to perform all monomial insertions, independently of the amount of minimisation taking place. This is taken up in Lem. \ref{nextmax}, Lem. \ref{CE-CountingSort} and Cor. \ref{embedded}. 

\begin{lem}
The cost to perform Next-Max-Order is $\theta(1)$ if buffer $A_1$ of Funnel Heap is non-empty. Else, the cost to perform Next-Max-Order accounts for the cost to Call one ensuing Extract-Max.
\label{nextmax}
\end{lem}
\begin{pf}
If buffer $A_1$ is non-empty, Next-Max-Order returns the maximum over all elements residing in the insertion buffer $S_{0,1}$ and $A_1$. Else, Next-Max-Order will trigger a FILL on buffer $A_1$ (see Sec. \ref{background}). Merely revealing the maximum element, however, does not alter the physical constituency of $A_1$. Hence, this buffer can be queried again with respect to the maximum residing in it, when the first Extract-Max to be encountered after the call to Next-Max-Order is issued. This can also be done without the need for FILL This completes the proof.
\end{pf}

\begin{lem}
Consider an invocation of Alg. FH-RANK during the $k$'th lifting step. If $k \in \theta(n)$, then sorting polynomial pairs on their total order in Step 2 of FH-RANK requires $\theta(k)$ work and $\theta((M+k)/B)$ cache misses. Else, we have $k \in O(n)$ but $k \notin \Omega(n)$, for which this step requires $O(k \log k)$ work and $O(\frac{k}{B} \log_{M/B} k)$ cache misses.
\label{CE-CountingSort}
\end{lem}
\begin{pf}
Collecting the total order of products in $\{(g_i\cdot h_j)\}_{k =1,\ldots,n}$ using a forward and backward scan of the arrays $Ord_g$ and $Ord_h$ respectively requires only cache miss operations, whose total cost amounts to $\theta(k/B)$. We now address the costs for sorting.

{\bf Case 1:} Using Counting Sort, the number of records to be sorted is equal to
$k-1$. Since each record represents the total degree of a monomial product, it is then less than or equal to $n-k$, and so, the work of counting sort is
$\theta(k + (n - k)) = \theta(n)$. Using a cache efficient variant of counting sort attains a cache complexity nearly
linear in the number of records as well, and is equal to
$\theta(\frac{n+M}{B})$ cache misses (see \cite{Moreno14}). 
When $k = \theta(n)$, the space, work, and cache complexity of Counting Sort simplify to those as stated in the
Lemma above. 

{\bf Case 2:} Here, we know that $k \notin \theta(n)$. But $k \leq n$, so we must have $k \in O(n)$ but $k \notin \Omega(n)$. Here, counting sort is no longer
linear in the number of records being sorted. Using any of the comparison-based, cache efficient sorting algorithms
requires $O(k)$ space, optimal $O(k \log k)$ work, and optimal 
$O(\frac{k}{B}\log_{M/B} k)$ cache misses (See \citep{FLPR99}, for example). 
\end{pf}

In the following Corollary, we conclude that the cost for sorting and pre-fetching the maximal order can be embedded in the asymptotic costs for performing all monomial products comprising $S_k$, independently of the amount of minimisation exerted onto Funnel Heap. 
\begin{cor}
Assume the sparse distributed representation for polynomials. Let $\left \vert T \right \vert_{\min}$ denote the size of Funnel Heap following the minimisation incurred by Alg. FH-RANK in the $k$'th lifting step, $\bar{g} = \max\{\#g_i\}_{i = 1}^{k}$ and $\bar{h} = \max \{\#h_j\}_{j = 1}^{k}$. Assume further that the cache size $M$ satisfies $M \in O(n)$, and, if $k \in O(n)$ but $k \notin \Omega(n)$, that $\bar{g}\bar{h} = \Omega(\log k)$. In the worst case analysis when each polynomial multiplication $g_ih_j$ is sparse, the cost to sort all polynomial products $\{(g_i\cdot h_j)\}$ on their total order and to issue all calls to NEXT-MAX-ORDER can be embedded in the cost for performing all insertions into Funnel Heap.  
\label{embedded}
\end{cor}
\begin{pf}
The gist of the proof lies in deriving bounds on the costs for the pre-processing phase that do not depend on $\left \vert T_{\min} \right \vert$. By Lem. \ref{nextmax}, each call to NEXT-MAX-ORDER accounts for the cost of the ensuing Extract-Max. 

Performing all $\theta \left (k \bar{g}\bar{h}\right)$ monomial extractions and insertions into Funnel Heap requires $O \left(k \bar{g}\bar{h} \log \left \vert T \right \vert_{\min} \right)$ work and amortised $O \left(k \bar{g}\bar{h} \frac{1}{B} \log_{M/B}  \left \vert T \right \vert_{\min} \right)$ cache misses. From Lem. \ref{CE-CountingSort}, if $k \in \theta(n)$, we know that cache friendly counting sort requires $\theta(k)$ work and $\theta((M+k)/B)$ cache misses, and for these costs to be embedded in their respective counterparts, we require 
\begin{equation}
k \in O\left (k \bar{g}\bar{h} \log \left \vert T \right \vert_{\min} \right)
\label{emb1}
\end{equation} 
and
\begin{equation}
\frac{(M+k)}{B} \in O\left (k \bar{g}\bar{h} \frac{1}{B} \log_{M/B}\left \vert T \right \vert_{\min}  \right).
\label{emb2}
\end{equation}
The requirement in (\ref{emb1}) trivially holds. For (\ref{emb2}), the assumption that $M \in O(n)$ when $k \in \theta(n)$ leads to $(M+k)/B = O(k/B) \in O\left (k \bar{g}\bar{h} \frac{1}{B} \log_{M/B}\left \vert T \right \vert_{\min}  \right)$.

When $k \in O(n)$ but $k \notin \Omega(n)$, the work of sorting ought to satisfy 
\begin{equation*}
k \log k \in O \left (k \bar{g}\bar{h} \log \left \vert T \right \vert_{\min} \right),
\end{equation*}
or 
\begin{equation*}
\log k \in O \left (\bar{g}\bar{h} \log \left \vert T \right \vert_{\min} \right),
\end{equation*}
which is attained if each monomial product $g_ih_j$, known to be sparse and having $O(\bar{g}\bar{h})$ non-zero terms, also has at least $\Omega(\log k)$ non-zero terms. This is also a very realistic assumption since $k$ is sufficiently small in this branch of the proof. Using this same requirement, we also get that the cache complexity required by sorting is embedded by that to perform all monomial insertions and extractions, as we can see from:
\begin{equation*}
\frac{k}{B}\log_{M/B} k \in O \left(k \bar{g}\bar{h} \frac{1}{B} \log_{M/B}  \left \vert T \right \vert_{\min} \right).
\end{equation*}
\end{pf}
\begin{rem}
The condition $M \in O(n)$ entails that the input polynomial has sufficiently high degree with respect to the size of in-core memory. This is a very reasonable requirement at scale bearing contemporary cache sizes.
To require that $\log k \in O(\bar{g}\bar{h})$ when $k \in O(n)$ but $k \notin \Omega(n)$ means that the sparsest polynomial product encountered in the $k$'th lifting step has $\Omega(\log k)$ non-zero terms. But any such product has degree at most $n-k$, which means that for significantly small iteration indices $k$, the polynomial products arising tend to be of significantly large degrees. The lower bound requirement on the sparsity of polynomial products indicated by $\bar{g}\bar{h} \in \Omega(\log k)$ is thus very permissive.
\end{rem}

\section{Experimental Results}
\label{experimental}

We implement all algorithms in C++ and compile our code using g++ version 4.4.6 20120305 with optimization level -O3. We run the experiments on an Intel(R) Xeon(R) CPU E5645 with 43GB of RAM, 12MB in L3 cache, and 256KB in L2 cache. To record cache misses, we use the STXXL library in Sec. \ref{benchmark} and the profiler tool perf in Sec. \ref{fhrank} and \ref{merging-tests}. 

\subsection{Funnel Heap Benchmarks: Performance at Scale}
\label{benchmark}

In this section we present a preamble where we benchmark Funnel Heap against Binary Heap for performing a generic sequence of priority queue operations outside the scope of Hensel lifting. The only available benchmarking appears in the unpublished work of \citep{SC08}\footnote{The authors of the current manuscript were unable to locate any standardised implementations of funnel heap available for public use.}. The specific goal of this section is to reproduce the conclusions derived in \citep{SC08} on our own machines and to reveal the cut-off line when Funnel Heap is able to beat Binary Heap. A careful examination is required before one is able to witness the performance predicted by the asymptotic analysis at large scale. This is because Funnel Heap performs more computations than Binary Heap, making it expensive to use at small scale, when the cost to perform memory accesses does not dominate performance. In line with \citep{SC08}, we simulate out of core behaviour by constraining the RAM of our machine to 16MB through the use of STXXL vectors (see STXXL version 1.3.1 and \citep{stxxl}). In this case, we force both heaps to store part of their structure on disk despite that the input suites are not too large. We generate a list of random integers and perform the following sequence of insertions and extractions onto the queues:
\begin{itemize}
  \item $N$ elements are pushed into the heap
  \item $N/2$ elements are popped off the heap
  \item $N/2$ elements are pushed into the heap
  \item $N$ elements are popped off the heap
\end{itemize}
In Table \ref{gpq} below, we present an account of the overall runtime as well as cache misses incurred by each of the two heaps. The term ``Max Capacity'' denotes the largest number of elements each heap occupies at any one point in time. In the first three rows, Funnel Heap loses out to Binary Heap in terms of overall runtime. For this particular range, the input is too small, and Funnel Heap's performance is computation bound, as it attempts to maintain its structure by calling FILL and SWEEP. Once Binary Heap grows out-of-core, however, its runtime becomes memory-bound and performance deteriorates significantly. This point in time is obtained when Binary Heap contains about $N = 4\times 10^6$ elements of size four bytes each, the expected cutoff line representing the customised size of RAM. Beyond that point, both heaps start swapping to disk and the cache complexity begins to dominate the computation cost. Funnel Heap now beats Binary Heap by orders of magnitude, as predicted by the asymptotic analysis. 

\begin{table}[htbp]
  \caption{Generic Priority Queue Operations}
\begin{center}
  \begin{tabular}{|c|c|c|c|} \hline
    Heap Type & Max Capacity &  Cache Misses &  Runtime \\
    &  & &  (s) \\ \hline
    Binary & 65,536 &  46,247 & 2.36 \\
    Binary & 262,144 &  220,761 & 9.76 \\
    Binary & 524,288 & 370,777  & 22.18 \\
    Binary & 1,048,576 & 987,864 & 46.49 \\
    Binary & 2,097,152 & 4,428,548 & 98.05 \\
    Binary & 4,194,304  & 16,364,635  & 206.67 \\
    Binary & 8,388,608 & 647,576,728 & 49,021.72 \\
    Binary & 16,777,216  & 3,977,883,205  & 337,626.06 \\ \hline
    Funnel & 65,536 & 142,839  & 2.85 \\
    Funnel & 262,144  & 291,070  & 10.91 \\
    Funnel & 524,288 & 779,126  & 26.21 \\
    Funnel & 1,048,576 & 1,572,436  & 55.89 \\
    Funnel & 2,097,152 & 3,319,421 & 127.17 \\
    Funnel & 4,194,304 & 5,073,462 & 269.02 \\
    Funnel & 8,388,608 & 13,639,451  & 558.49 \\
    Funnel & 16,777,216  & 23,786,264 & 1,234.35 \\ \hline
  \end{tabular}
\label{gpq}
\end{center}
\end{table}

The results in this section should be construed in the following sense. The input suite to our polynomial factorisation experiments below does not attain a level of growth sufficient to solicit out-of core behaviour. As such, any significant improvements in performance shown hereafter can only be attributed to the optimisations incurred onto Funnel Heap, particularly, batched chaining and the optimised sequence of insertions, but not to Funnel Heap alone. On the other hand, we do expect Funnel Heap to outperform Binary Heap considerably, without any of those mentioned optimisations. This will remain applicable even when tackling a single polynomial multiplication or division -- and not just sums of products -- once the input data grows sufficiently large.

\subsection{Performance of FH-RANK}
\label{fhrank}


This section is dedicated to the performance of FH-RANK. We start off with a summary of relevant results from \citep{AEG15}, where we observe all of the following. Both overlapping implementations PQ-HL$^{\mathcal{B}}$ and FH-HL are always significantly faster than SER-HL, despite that the Newton polygons treated there are extremely sparse. Even when the polygon has a few edges, the polytope method remains susceptible to fluctuations in the sparsity of the intermediary polynomials, which is attributed to expression swell. Both PQ-HL$^{\mathcal{B}}$ and FH-HL are also always faster than Magma 2.18-7, whose built-in function for factoring bivariate polynomials relies on the standard algorithms in \citep{BM97,vH02}. Starting with bivariate polynomials of total degree equal to $10,000$, and despite that the input polynomials are significantly sparse, Magma runs out of memory. Alg. FH-HL is always fastest, with a dramatic reduction in run-time over SER-HL which is largely attributed to substantial expression swell, and also over PQ-HL$^{\mathcal{B}}$, which is largely attributed to BATCHED-CHAIN.

Hereafter we address the performance of FH-RANK, by benchmarking it against all of PQ-HL$^{\mathcal{F}}$, PQ-HL$^{\mathcal{B}}$, PQ-HL$^{\mathcal{B}}$ with chainining (PQ-HL$^{\mathcal{B}}$-CHAIN), and FH-HL. We use the sparse distribued representation for encoding all polynomials. 
Our input suite consists of random bivariate polynomials over $\mathbb{F}_3$ that turn out to factor into two irreducibles, and that are extremely sparse. In several instances, we specifically generate random polynomials whose Newton polygon is the sparsest possible, consisting of the triangle $(0,n)$, $(n,0$), and $(0,0)$. For sorting the polynomial pairs on their total degree, we note that the input degrees we treat here fall within a certain range for which the standard GCC quicksort implementation is known to be highly competitive over cache efficient alternatives. This is established thoroughly by the algorithmic engineering study of \citep{BFV08} (see for example the experiments in Sec. 5). 

The results for each input polynomial tested are reported using a pair of tables. In all of the captions, the parameter $n$ denotes the total degree of the input polynomial, and $t$ denotes the total number of its non-zero terms. Our requirement for sparse input is to have $t \ll n^2$. The parameter $F$ corresponds to the total number of boundary factorisations attempted before the two irreducible factors are produced. The first table for each input polynomial provides an account of run-time in seconds as well as cache misses. 
The second and third ensuing tables show the number of SWEEPs called upon each link in the Funnel Heap used in Alg. PQ-HL$^{\mathcal{F}}$ and Alg. FH-RANK respectively. 
In those tables we also indicate the size of each link.

{\bf FH-RANK against all other variants:} Both FH-RANK and FH-HL are faster than all of PQ-HL$^{\mathcal{F}}$, PQ-HL$^{\mathcal{B}}$, and PQ-HL$^{\mathcal{B}}$-CHAIN, and they incur considerably less cache misses thanks to BATCHED-CHAIN. In turn, FH-RANK improves over FH-HL at a larger scale thanks to the optimised sequence of insertions. This is demonstrated by an order of magnitude reduction in time as well as cache misses over FH-HL. For all of the input polynomials tested, the second and third tables show that FH-RANK brings about an order of magnitude reducion in space, as demonstrated in the reduction of the number of links required, as well as the corresponding size of each link. Of significance also is the notable reduction in the number of sweeps to each link. The improvement in the amount of space required as well as the number of SWEEPs incurred explain the significant reduction in the run-time and cache misses associated with FH-RANK. In contrast, we note that FH-HL consumed the same amount of peak memory as PQ-HL$^{\mathcal{F}}$: as explained in \citep{AEG15}, FH-HL requires the same amount of space as FH, except that an asymptotically large amount of space shifts from being ``working space'' to ``auxiliary space'', which improves on runtime and rate of cache misses of FH-HL. 

{\bf The effect of chaining on Binary Heap:} In several instances, PQ-HL$^{\mathcal{B}}$-CHAIN incurs the same order of cache misses as PQ-HL$^{\mathcal{B}}$, and on a few occasions it is actually slower. This demonstrates that chaining is not consistently efficient when employing Binary Heap. In \citep{AEG15}, we elaborate on the reasons behind this behaviour. For example, each insertion into the linked list denoting the chain incurs a random miss. Also, a single search for a replica to be chained may very well require traversing the entire heap of size $N$, bringing the work and cache complexity for performing a single search to be that of traversal of $N$ elements. Summing up, neither temporal locality nor spatial locality are observed in PQ-HL$^{\mathcal{B}}$-CHAIN. 

{\bf Funnel Heap without any of the optimisations:} The results reported for PQ-HL$^{\mathcal{F}}$ are also not promising for the range of input degrees we had treated, specifically as taken against PQ-HL$^{\mathcal{B}}$ and PQ-HL$^{\mathcal{B}}$-CHAIN. Funnel Heap inherently performs more work, and incurs more cache misses for smaller levels of the memory hierarchy. The input polynomials treated here do not solicit access to disk. As a result, the extra work performed by Funnel Heap is not compensated for. Yet, it is evident from the benchmarks reported in Section \ref{benchmark}, that PQ-HL$^{\mathcal{F}}$ is set to outperform PQ-HL$^{\mathcal{B}}$ and PQ-HL$^{\mathcal{B}}$-CHAIN once the input is large enough to solicit disk swaps. At smaller scale, all the benefits observed are attributed solely to the techniques in BATCHED-CHAIN and optimising the sequence of insertions (FH-HL and FH-RANK respectively).

\subsubsection*{}

\begin{table}[!htb]
\caption{Input: $n = 2000$, $t \approx 10^6$, $F = 1$}
\vspace{1.5ex} \centering
\begin{tabular}{|c|c|c|} \hline
Heap Type &   Runtime & Cache \\
  &    (s) & Misses \\
  &   &   \\\hline
PQ-HL$^{\mathcal{B}}$ & $4.2''$ &  121,419 \\
PQ-HL$^{\mathcal{B}}$-Chain  & $7.46''$ &  147,172 \\
PQ-HL$^{\mathcal{F}}$  & $7.32''$  & 123,857\\
FH-HL  & $3.92''$ &  158,182 \\
FH-RANK & $4.62''$ &  $104,545$\\
\hline
\end{tabular}
\end{table}
\begin{table}[!htb]
\caption{Sweeps per link in PQ-HL$^{\mathcal{F}}$ -- $n = 2000$, $t \approx 10^6$, $F = 1$}
\vspace{1.5ex} \centering
\begin{tabular}{|c|c|c|c|c|c|c|} \hline
 Link 1 & Link 2 & Link 3 & Link 4 & Link 5 & Link 6 & Link 7 \\
 \tiny{1.78E-04 MB}&  \tiny{1.2E-03 MB}  &  \tiny{9.22E-03 MB} &  \tiny{0.1 MB} &  \tiny{2.5 MB}  & \tiny{20 MB}  &  \tiny{11E+03 MB}  \\\hline
102,334	& 40,960 &	9,300	& 437&	0 &	0 & 0  \\
\hline
\end{tabular}
\end{table}
\begin{table}[!htb]
\caption{Sweeps per link in FH-RANK -- $n = 2000$, $t \approx 10^6$, $F = 1$}
\vspace{1.5ex} \centering
\begin{tabular}{|c|c|c|c|c|c|c|} \hline
 Link 1 & Link 2 & Link 3 & Link 4 & Link 5 & Link 6 & Link 7 \\
 \tiny{1.78E-04 MB}&  \tiny{1.2E-03 MB}  &  \tiny{9.22E-03 MB} &  \tiny{0.1 MB} &  \tiny{2.5 MB}  & \tiny{20 MB}  &  \tiny{11E+03 MB}  \\\hline
 91,723 & 36,889& 	8,305	& 253	& 0	& 0& 0 \\
\hline
\end{tabular}
\end{table}

\subsubsection*{}

\begin{table}[!htb]
\caption{Input: $n = 4000$, $t \approx 2 \times 10^2$, $F = 8$}
\vspace{1.5ex} \centering
\begin{tabular}{|c|c|c|} \hline
Heap Type & Runtime & Cache \\
  &     (s) & Misses \\
  &   &   \\\hline
PQ-HL$^{\mathcal{B}}$ & $32''$ & 86,068,900 \\
PQ-HL$^{\mathcal{B}}$-Chain  & $17.91''$ &  85,762,222\\
PQ-HL$^{\mathcal{F}}$ & $115.94''$ & 87,130,225 \\
FH-HL  & $16.048''$ & 85,860,379 \\
FH-RANK & $15.73''$ & $14,310,063$\\
\hline
\end{tabular}
\end{table}

\begin{table}[!htb]
\caption{Sweeps per link in PQ-HL$^{\mathcal{F}}$ -- $n = 4000$, $t \approx 2 \times 10^2$, $F = 8$}
\vspace{1.5ex} \centering
\begin{tabular}{|c|c|c|c|c|c|c|} \hline
 Link 1 & Link 2 & Link 3 & Link 4 & Link 5 & Link 6 & Link 7 \\
 \tiny{1.78E-04 MB}&  \tiny{1.2E-03 MB}  &  \tiny{9.22E-03 MB} &  \tiny{0.1 MB} &  \tiny{2.5 MB}  & \tiny{20 MB}  &  \tiny{11E+03 MB}  \\\hline
33,285,346	& 13,301,964& 	2,953,392& 	281,734	& 0& 	0& 0\\
\hline
\end{tabular}
\end{table}

\begin{table}[!htb]
\caption{Sweeps per link in FH-RANK -- $n = 4000$, $t \approx 2 \times 10^2$, $F = 8$}
\vspace{1.5ex} \centering
\begin{tabular}{|c|c|c|c|c|c|c|} \hline
 Link 1 & Link 2 & Link 3 & Link 4 & Link 5 & Link 6 & Link 7 \\
 \tiny{1.78E-04 MB}&  \tiny{1.2E-03 MB}  &  \tiny{9.22E-03 MB} &  \tiny{0.1 MB} &  \tiny{2.5 MB}  & \tiny{20 MB}  &  \tiny{11E+03 MB}  \\\hline
112,525& 	25,860&	70&	2&	0&	0 & 0\\
\hline
\end{tabular}
\end{table}

\newpage

\subsubsection*{}

\begin{table}[!htb]
\caption{Input: $n = 5000$, $t \approx 2 \times 10^2$, $F = 5$}
\vspace{1.5ex} \centering
\begin{tabular}{|c|c|c|} \hline
Heap Type &   Runtime & Cache \\
  &    (s) & Misses \\
  &   &   \\\hline
PQ-HL$^{\mathcal{B}}$  & $64''$ & 49,240,337  \\
PQ-HL$^{\mathcal{B}}$-Chain  & $39.4''$  &  49,483,164\\
PQ-HL$^{\mathcal{F}}$& $193.06$ & 50,129,826 \\
FH-HL  & $32.05''$ &  48,999,253\\
FH-RANK & $27.55''$ & $5,389,917$ \\
\hline
\end{tabular}
\end{table}

\begin{table}[!htb]
\caption{Sweeps per link in PQ-HL$^{\mathcal{F}}$ -- $n = 5000$, $t \approx 2 \times 10^2$, $F = 5$}
\vspace{1.5ex} \centering
\begin{tabular}{|c|c|c|c|c|c|c|} \hline
 Link 1 & Link 2 & Link 3 & Link 4 & Link 5 & Link 6 & Link 7 \\
 \tiny{1.78E-04 MB}&  \tiny{1.2E-03 MB}  &  \tiny{9.22E-03 MB} &  \tiny{0.1 MB} &  \tiny{2.5 MB}  & \tiny{20 MB}  &  \tiny{11E+03 MB}  \\\hline
20,309,315 &	8,119,178&	1,802,960&	195,408	& 820& 	0& 0\\
\hline
\end{tabular}
\end{table}

\begin{table}[!htb]
\caption{Sweeps per link in FH-RANK -- $n = 5000$, $t \approx 2 \times 10^2$, $F = 5$}
\vspace{1.5ex} \centering
\begin{tabular}{|c|c|c|c|c|c|c|} \hline
 Link 1 & Link 2 & Link 3 & Link 4 & Link 5 & Link 6 & Link 7 \\
 \tiny{1.78E-04 MB}&  \tiny{1.2E-03 MB}  &  \tiny{9.22E-03 MB} &  \tiny{0.1 MB} &  \tiny{2.5 MB}  & \tiny{20 MB}  &  \tiny{11E+03 MB}  \\\hline
109,553	& 30,540&	574&	9&	0&	0&0\\
\hline
\end{tabular}
\end{table}

\newpage

\subsubsection*{}

\begin{table}[!htb]
\caption{Input: $n = 6000$, $t \approx 8 \times 10^2$, $F = 3$}
\vspace{1.5ex} \centering
\begin{tabular}{|c|c|c|} \hline
Heap Type &  Runtime & Cache \\
  &     (s) & Misses \\
  &   &   \\\hline
PQ-HL$^{\mathcal{B}}$ & $810''$ & 2,953,189,933 \\
PQ-HL$^{\mathcal{B}}$-Chain  & $372.24''$ & 2,976,047,883 \\
PQ-HL$^{\mathcal{F}}$& $2754''$ &  3,005,649,954 \\
FH-HL  &  $300.33''$ & 2,933,746,216 \\
FH-RANK & $241.64''$ & $586,749,243$ \\
\hline
\end{tabular}
\end{table}

\begin{table}[!htb]
\caption{Sweeps per link in PQ-HL$^{\mathcal{F}}$ -- $n = 6000$, $t \approx 8 \times 10^2$, $F = 3$}
\vspace{1.5ex} \centering
\begin{tabular}{|c|c|c|c|c|c|c|} \hline
 Link 1 & Link 2 & Link 3 & Link 4 & Link 5 & Link 6 & Link 7 \\
 \tiny{1.78E-04 MB}&  \tiny{1.2E-03 MB}  &  \tiny{9.22E-03 MB} &  \tiny{0.1 MB} &  \tiny{2.5 MB}  & \tiny{20 MB}  &  \tiny{11E+03 MB}  \\\hline
 1,399,287,606& 	559,616,518& 	124,332,749& 	14,567,724& 	626,862& 	0& 0\\
\hline
\end{tabular}
\end{table}

\begin{table}[!htb]
\caption{Sweeps per link in FH-RANK -- $n = 6000$, $t \approx 8 \times 10^2$, $F = 3$}
\vspace{1.5ex} \centering
\begin{tabular}{|c|c|c|c|c|c|c|} \hline
 Link 1 & Link 2 & Link 3 & Link 4 & Link 5 & Link 6 & Link 7 \\
 \tiny{1.78E-04 MB}&  \tiny{1.2E-03 MB}  &  \tiny{9.22E-03 MB} &  \tiny{0.1 MB} &  \tiny{2.5 MB}  & \tiny{20 MB}  &  \tiny{11E+03 MB}  \\\hline
5,496,504& 	1,871,555& 	82,201& 	436& 	0& 	0& 0\\
\hline
\end{tabular}
\end{table}

\subsubsection*{}

\begin{table}[!htb]
\caption{Input:  $n = 10,000$, $t \approx 3 \times 10^5$, $F = 7$}
\vspace{1.5ex} \centering
\begin{tabular}{|c|c|c|} \hline
Heap Type &  Runtime & Cache \\
  &     (s) & Misses \\
  &   &   \\\hline
PQ-HL$^{\mathcal{B}}$  & $17$ hrs $19'$ $15''$ & 108,753,910,699	 \\
PQ-HL$^{\mathcal{B}}$-Chain  & $9$ hrs $39'$ $33''$ & 378,734,677,525 \\
PQ-HL$^{\mathcal{F}}$& $51$ hrs $18'$ $11''$ & 154,290,119,919 \\
FH-HL & $6$ hrs $56'$ $40''$  & 75,552,675,674 \\
FH-RANK & $4$ hrs $49'$ $24''$& $1,662,290,864$ \\
\hline
\end{tabular}
\end{table}

\begin{table}[!htb]
\caption{Sweeps per link in PQ-HL$^{\mathcal{F}}$ -- $n = 10,000$, $t \approx 3 \times 10^5$, $F = 7$}
\vspace{1.5ex} \centering
\begin{tabular}{|c|c|c|c|c|c|c|} \hline
 Link 1 & Link 2 & Link 3 & Link 4 & Link 5 & Link 6 & Link 7 \\
 \tiny{1.78E-04 MB}&  \tiny{1.2E-03 MB}  &  \tiny{9.22E-03 MB} &  \tiny{0.1 MB} &  \tiny{2.5 MB}  & \tiny{20 MB}  &  \tiny{11E+03 MB}  \\\hline
36,352,370,146 & 	14,526,199,363 & 	3,217,623,808& 	374,739,688& 	22,680,945& 	60,352 & 0\\
\hline
\end{tabular}
\end{table}

\begin{table}[!htb]
\caption{Sweeps per link in FH-RANK -- $n = 10,000$, $t \approx 3 \times 10^5$, $F = 7$}
\vspace{1.5ex} \centering
\begin{tabular}{|c|c|c|c|c|c|c|} \hline
 Link 1 & Link 2 & Link 3 & Link 4 & Link 5 & Link 6 & Link 7 \\
 \tiny{1.78E-04 MB}&  \tiny{1.2E-03 MB}  &  \tiny{9.22E-03 MB} &  \tiny{0.1 MB} &  \tiny{2.5 MB}  & \tiny{20 MB}  &  \tiny{11E+03 MB}  \\\hline
153,656,353& 	56,411,062& 	2,672,402& 	17,629& 	0& 	0 & 0\\
\hline
\end{tabular}
\end{table}

\subsubsection*{}

\begin{table}[!htb]
\caption{Input: $n = 13,000$, $t \approx 2 \times 10^6$, $F = 2$}
\vspace{1.5ex} \centering
\begin{tabular}{|c|c|c|} \hline
Heap Type &  Runtime & Cache \\
  &    (s) & Misses \\
  &   &   \\\hline
PQ-HL$^{\mathcal{B}}$  &  511.93 &   5,107,869,507 \\
PQ-HL$^{\mathcal{B}}$-Chain   & 7,799.34 & 5,307,782,843 \\
PQ-HL$^{\mathcal{F}}$ & 7,650.19 & 3,110,577,104\\
FH-HL  & 179.602 & 845,684,286 \\
FH-RANK & $123.33$ & $56,378,952$ \\
\hline
\end{tabular}
\end{table}

\begin{table}[!htb]
\caption{Sweeps per link in PQ-HL$^{\mathcal{F}}$ -- $n = 13,000$, $t \approx 2 \times 10^6$, $F = 2$}
\vspace{1.5ex} \centering
\begin{tabular}{|c|c|c|c|c|c|c|} \hline
 Link 1 & Link 2 & Link 3 & Link 4 & Link 5 & Link 6 & Link 7 \\
 \tiny{1.78E-04 MB}&  \tiny{1.2E-03 MB}  &  \tiny{9.22E-03 MB} &  \tiny{0.1 MB} &  \tiny{2.5 MB}  & \tiny{20 MB}  &  \tiny{11E+03 MB}  \\\hline
105,139,144 & 42,055,658 & 9,345,701 & 1,099,494 & 66,636 & 2,066 & 16 \\
\hline
\end{tabular}
\end{table}

\begin{table}[!htb]
\caption{Sweeps per link in FH-RANK -- $n = 13,000$, $t \approx 2 \times 10^6$, $F = 2$}
\vspace{1.5ex} \centering
\begin{tabular}{|c|c|c|c|c|c|c|} \hline
 Link 1 & Link 2 & Link 3 & Link 4 & Link 5 & Link 6 & Link 7 \\
 \tiny{1.78E-04 MB}&  \tiny{1.2E-03 MB}  &  \tiny{9.22E-03 MB} &  \tiny{0.1 MB} &  \tiny{2.5 MB}  & \tiny{20 MB}  &  \tiny{11E+03 MB}  \\\hline
29,000,849 & 11,605,823 & 2,686,025 & 206,611 & 1,943 & 1 & 0 \\
\hline
\end{tabular}
\end{table}

\newpage

\subsection{Funnel Heap versus the $k$-merger}
\label{merging-tests}

In this section we gather further insight into the bottleneck associated with the irregularity in computations as a result of the varying density of the intermediary output arising during Hensel lifting. In \citep{AEG14}, we observed that if space is not a concern and one is willing to store all polynomial products during each Hensel lifting step before their final merge into $S_k$, the only cache oblivious competitor at scale for the global priority queue approach would be the following scenario. One would perform each polynomial multplication separately using a dedicated (local) priority queue, followed by merging of all the polynomial products using the static and cache oblivious $k$-merger of \citep{FLPR99}. Local priority queues tackling one polynomial multiplication at a time are more likely to reside in cache when the multiplication is sparse. As such, a data structure suited for internal memory such as a MAX-heap should be used, since it performs less work than any external memory implementation. The goal of this section, however, is to demonstrate that even in this scenario, the $k$-merger still fails to achieve its optimal (and amortised) work and cache complexity, since the streams to be merged are of varying density. This is in contrast to its typical mode of usage in merge-based sorting algorithms where the streams tend to be of equal size. This section provides an empirical proof of concept that overlapping arithmetic using a global funnel heap is not only cache-oblivious, but is further guaranteed to attain optimal performance, bearing the irregularity in computation. An interesting conclusion we draw is that, despite that the $k$-merger incurs optimal work for merging a given set of elements, it is beaten by funnel heap, albeit at a higher work complexity, once the $k$-merger fails to amortise its cache complexity in the presence of irregular input.

To this end, we demonstrate by merging streams of random integers, and we simulate three scenarios depending on the density of streams. In Table \ref{ksq-k}, we process $k^2$ streams each containing $k$ elements, in Table \ref{k-k} we process $k$ streams with $k$ elements each, and in Table \ref{k-ksq}, we process $k$ streams with $k^2$ elements each. 
In addition to cache misses and total execution time, we record the total number of integer comparisons required by merging sub-routines, including the merging that is required by the SWEEP function in Funnel Heap. 

{\it On integer comparisons}: In all three tables, Funnel Heap
does more integer comparisons than the $k$-merger, something we
attribute to the cost of its SWEEP function. Despite that it incurs
the least number of comparisons, the $k$-merger lags behind in cache miss rate and then finally overall execution time.

{\it On cache misses}: In all three tables, Funnel Heap
terminates using significantly lower cache misses than the
$k$-merger. We note, however, that the rate of improvement of Funnel
Heap in Table \ref{k-ksq} is less than what is observed for Tables
\ref{ksq-k} and \ref{k-k}, which is to be expected, since the
$k$-merger is now able to produce $k^3$ elements at the end of its invocation.
This is in line with the $k$-merger's cache complexity analysis, by which we know that
the amortised cache complexity is met so long as the output buffer produces $k^3$ elements. 

{\it On overall runtime}: In all three categories, Funnel Heap terminates 
faster than the single, fixed size $k$-merger. In Tables \ref{ksq-k}
and \ref{k-k}, Funnel Heap attains a significantly lower cache miss rate that justifies its fast execution time. In Table \ref{k-ksq},
and despite that the $k$-merger catches up by performing its best cache misse rate as opposed to Tables  \ref{ksq-k}
and \ref{k-k}, it still lags behind Funnel Heap in
total execution time.

\begin{table}[!htbp]
\caption{$k^2$ streams of $k$ elements each}
\begin{center}
\begin{tabular}{|c|c|c|c|c|c|c|} \hline
Merge Type & Streams & Input Per & Insertions, & Cache & Comparison & Runtime  \\
  &   & Stream & Extractions & Misses & Count & (s)  \\\hline
FunnelHeap & 64 & 8 & 512 & 18,495 & 5,900 & $0.005''$ \\
FunnelHeap & 128 & 11 & 1,408 & 15,603 & 20,217 & $0.010''$ \\
FunnelHeap & 256 & 16 & 4,096 & 21,100 & 63,524 & $0.019''$ \\
FunnelHeap & 512 & 22 & 11,264 & 25,771 & 209,359 & $0.049''$ \\
Kmerger & 64 & 8 & 512 & 367,006 & 2,937 & $0.424''$ \\
Kmerger & 128 & 11 & 1,408 & 1,171,682 & 9,596 & $2.059''$ \\
Kmerger & 256 & 16 & 4,096 & 8,206,508 & 32,248 & $21.963''$ \\
Kmerger & 512 & 22 & 11,264 & 74,409,763 & 100,410 & $132.328''$ \\
\hline
\end{tabular}
\end{center}
\label{ksq-k}
\end{table}
\begin{table}[!htbp]
\caption{$k$ streams with $k$ elements each}
\begin{center}
\begin{tabular}{|c|c|c|c|c|c|c|} \hline
Merge Type & Streams & Input Per & Insertions, & Cache & Comparison & Runtime \\
  &   & Stream & Extractions & Misses & Count & (s)  \\\hline
FunnelHeap & 64 & 64 & 4,096 & 13,482 & 59,641 & $0.020''$ \\
FunnelHeap & 128 & 128 & 16,384 & 19,135 & 290,088 & $0.053''$ \\
FunnelHeap & 256 & 256 & 65,536 & 63,775 & 1,378,606 & $0.288''$ \\
FunnelHeap & 512 & 512 & 262,144 & 129,492 & 6,394,298 & $0.941''$ \\
Kmerger & 64 & 64 & 4,096 & 354,527 & 24,451 & $0.598''$ \\
Kmerger & 128 & 128 & 16,384 & 1,935,724 & 114,412 & $2.242''$ \\
Kmerger & 256 & 256 & 65,536 & 6,975,425 & 523,766 & $17.972''$ \\
Kmerger & 512 & 512 & 262,144 & 52,426,417 & 2,358,303 & $130.263''$ \\
\hline
\end{tabular}
\end{center}
\label{k-k}
\end{table}
\begin{table}[!htbp]
\caption{$k$ streams with $k^2$ elements each}
\begin{center}
\begin{tabular}{|c|c|c|c|c|c|c|} \hline
Merge Type & Streams & Input Per & Insertions, & Cache & Comparison & Runtime \\
  &   & Stream & Extractions & Misses & Count & (s)  \\\hline
FunnelHeap & 64 & 4,096 & 262,144 & 169,123 & 6,270,855 & $0.881''$ \\
FunnelHeap & 128 & 16,384 & 2,097,152 & 1,693,942 & 63,857,189 & $12.121''$ \\
FunnelHeap & 256 & 65,536 & 16,777,216 & 17,161,816 & 535,670,619 & $180.296''$ \\
FunnelHeap & 512 & 262,144 & 134,217,728 & 261,091,499 & 4,752,491,629 & $3,169.407''$ \\
Kmerger & 64 & 4,096 & 262,144 & 433,473 & 1,572,732 & $0.854''$ \\
Kmerger & 128 & 16,384 & 2,097,152 & 5,138,297 & 14,679,774 & $44.939''$ \\
Kmerger & 256 & 65,536 & 16,777,216 & 39,108,825 & 134,217,167 & $560.544''$ \\
Kmerger & 512 & 262,144 & 134,217,728 & 265,845,601 & 1,207,958,504 & $3,919.140''$ \\
\hline
\end{tabular}
\end{center}
\label{k-ksq}
\end{table}

\newpage

\section{Conclusion}

In this paper we presented a comprehensive design and analysis that extends the work in \citep{AEG14} and \citep{AEG15}. Alg. FH-RANK exploits all the features of Funnel Heap for implementing sums of products arising in Hensel lifting of the polytope method, when polynomials are in sparse distributed representation. Those features involve a batched mechanism for chaining replicas as well as optimising on the sequence of insertions and extractions in order to minimise the size of the priority queue as well as the work and cache complexity. The competitive asymptotics are validated by empirical results, which, in addition to asserting the high efficieny of FH-RANK whether or not data fits in in-core memory, help us derive two other main conclusions. Firstly, we confirm that at a large scale, all polynomial arithmetic employing a priority queue will benefit substantially from using Funnel Heap over Binary Heap, even without the proposed mechanisms for chaining and/or optimising the sequence of insertions/extractions. Secondly, Funnel Heap is confirmed to be superior in practice as a merger when tested against the provably optimal $k$-merger structure, despite having a higher work complexity. This is attributed to its ability to adapt to merging input streams of fluctuating density, which in turn, makes Funnel Heap ideal for performing polynomial arithmetic in the sparse distributed representation, where such fluctuation affects overall performance. This supports our argument that one should resort to the overlapping approach using a single priority queue, as opposed to handling each of the the local multiplications separately using a local priority queue, to be followed by additive merging of all polynomial streams. This conclusion remains valid whether or not expression swell is taking place.

\section{Acknowledgments}
We thank the Lebanese National Council for Scientific Research and
the University Research Board -- American University of Beirut, for
supporting this work.

\section*{References}



\end{document}